\documentclass[reprint,aps,prl,superscriptaddress,showpacs,preprintnumbers]{revtex4-1}

\usepackage[usenames,dvipsnames]{color}
\usepackage{amssymb}
\usepackage{graphicx}
\usepackage{cancel}
\usepackage{amsmath}
\usepackage{braket}

\begin{document}


\title{Slow Thermalization Between a Lattice and Free Bose Gas}
\author{David C. McKay}
    \altaffiliation[Now at:]{James Franck Institute, University of Chicago, 929 E 57th Street, Chicago, IL 60637}
\author{Carolyn Meldgin}
\author{David Chen}
\author{Brian DeMarco}
\affiliation{Department of Physics, University of Illinois, 1110 W Green St., Urbana, IL 61801}
\date{\today}

\begin{abstract}
Using a 3D spin-dependent optical lattice, we study thermalization and energy exchange between two ultracold Bose gases, one of which is strongly correlated and bound to the lattice and another that is free from the lattice potential.  Disruption of inter-species thermalization is revealed through measurements of condensate fraction after the lattice is superimposed on the parabolic confining potential.  By selectively heating the lattice-bound species and measuring the rate of heat transfer to the free state, suppression of energy exchange is observed.  Comparison with a Fermi's golden rule prediction confirms that this effect is caused by a dispersion mismatch that reduces the phase space available for elastic collisions.  This result has critical implications for methods proposed to cool strongly correlated lattice gases.
\end{abstract}


\maketitle

Ultracold atom gases trapped in optical lattices are an ideal system for exploring strongly correlated quantum matter.  Initial work in this area focused on investigating equilibrium phases of Hubbard models, including the realization of the superfluid-to-Mott-insulator phase transition for bosons \cite{greiner:2002} and the metal-to-Mott-insulator transition for fermions \cite{jordens:2008,schneider:2008}.  We explore a new frontier in this field---out-of-equilibrium dynamics---that presents outstanding questions to state-of-the-art numerical approaches and theory.  Recent results in this area include measurements of quantum quenches \cite{PhysRevLett.106.235304,Chenueau2012}, expansion dynamics \cite{Schneider2012}, and conduction \cite{Brantut31082012}.  These measurements provide insight into phenomena important to material applications, such as transport and diffusion.  Understanding strongly correlated dynamics is also critical to developing new techniques for cooling lattice gases.  Presently attainable temperatures in optical lattices are too high to observe many phases of interest, such as antiferromagnetism \cite{mckay:2011}.  New methods for cooling are therefore required to reach these low-entropy states.

Here we examine the dynamics most relevant to cooling---thermalization and energy exchange---in a unique strongly correlated system: a fully three-dimensional, species-specific optical lattice \cite{mckay:2010}.  We work with a mixture of two bosonic atomic species, one that is strongly correlated and bound to the lattice, and another weakly interacting state free from the lattice potential.  This type of lattice has been proposed as a method for realizing exotic bosonic and fermionic superfluid states \cite{PhysRevLett.90.100401,PhysRevA.70.033603}, a Kondo lattice model \cite{PhysRevA.82.053624,PhysRevA.81.051603}, and for quantum magnetism \cite{PhysRevLett.91.090402}.  The measurements we present here impact those proposals and suggestions to use this system as a platform for cooling and thermometry of strongly correlated lattice gases.  For cooling, a large number of free atoms may be used as a heat reservoir to directly absorb entropy from the lattice gas when it is compressed \cite{ho:2009} or to accept entropy via band decay processes \cite{griessner:2006}, while a small impurity of free atoms could be employed for thermometry \cite{mckay:2010}.


There are two methods for realizing species-specific lattices.  In one scheme, different atomic species or isotopes are employed, and the relative lattice potential depths are adjusted via the lattice laser wavelength \cite{leblanc:2007,catani:2009}. We use the other technique, which is to employ a mixture of atoms in different hyperfine (i.e., spin) states trapped in a spin-dependent potential \cite{deutsch:1998}. Spin-dependent lattices in 1D have been used to generate entanglement \cite{mandel:2003}, realize quantum walks \cite{karski:2009}, create atomic impurity disorder \cite{gadway:2011}, realize mixed-dimensional systems \cite{PhysRevLett.104.153202}, and to implement a matter-wave probe \cite{gadway:2012}. We have used a 3D spin-dependent lattice to create mixed superfluid and Mott insulator phases \cite{mckay:2010}.



\begin{figure}
\includegraphics[width=0.45\textwidth]{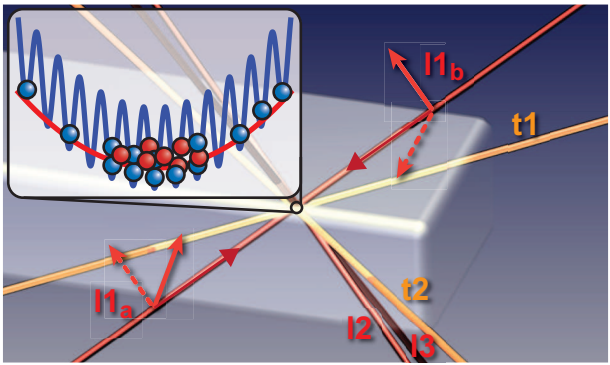}
\caption{(Color online) Schematic of the spin-dependent lattice potential. The $\Ket{1}$ atoms (blue) are lattice-bound, while the $\Ket{0}$ atoms (red) do not experience the lattice potential.  Confinement of both species is provided by a 1064~nm crossed-dipole trap (orange lines, t1 and t2). The spin-dependent lattice is formed by three pairs counter-propagating 790~nm laser beams (red lines, l1, l2, and l3).  Each pair consists of beams with orthogonal linear polarizations
(indicated by dashed arrows for $\text{l1}_\text{a}$ and $\text{l2}_\text{b}$). \label{fig:schematic}}
\end{figure}

We use the $\Ket{F=1,m_F=-1}$ (``$\Ket{1}$") and $\Ket{F=1,m_F=0}$ (``$\Ket{0}$") states of $^{87}$Rb atoms confined in a 1064~nm crossed-dipole trap; the full details of our apparatus and can be found in Ref.~\cite{mckay:2010}.  A mixture of $\Ket{1}$ and $\Ket{0}$ atoms is created from a Bose-Einstein condensate (BEC) of $\Ket{1}$ atoms via adiabatic rapid passage driven by a radio-frequency magnetic field.  This mixture is miscible in the dipole trap and stabilized against spin-changing collisions by a 10--30 gauss magnetic field \cite{supp}.  The atoms are transferred into a spin-dependent cubic lattice by slowly superimposing three orthogonal pairs of counter-propagating, linearly polarized $\lambda=790$~nm laser beams in a lin-$\perp$-lin configuration (Fig.~\ref{fig:schematic}). At this wavelength, the lattice potential depth is proportional to $\left|m_F\right|$ \cite{mckay:2010}.  Thus, the $\Ket{1}$ atoms are lattice-bound and realize the Bose-Hubbard (BH) model, and the $\Ket{0}$ atoms do not experience the lattice, as shown in Fig.~\ref{fig:schematic}.  Using a combination of direct imaging and microwave spectroscopy \cite{supp}, we confirm that the two species are spatially overlapped on the trap and lattice length scales for all the lattice potential depths explored in this work. We express the lattice potential depth $s$ in units of the recoil energy $E_R=2\hbar^2\pi^2/m\lambda^2$ (where $m$ is the atomic mass).

By measuring condensate fraction, we observe that turning on the lattice disrupts inter-species thermalization.  The lattice laser intensity is turned on over 50~ms, slower than the observed 30~ms inter-species thermalization time in the dipole trap \cite{mckay:2010}. The lattice modifies the dispersion of the lattice-bound atoms and reduces their kinetic energy, and therefore, if the lattice turn-on is isentropic, the $\Ket{1}$ atoms will cool.  The temperature of the $\Ket{0}$ atoms should likewise decrease if the species remain in thermal equilibrium, and thus condensate fraction $N_0/N$ for that species will increase.  We measure $N_0/N$ for each species using time-of-flight imaging as $s$ is varied. Separate images of each component are obtained after turning off the lattice and dipole trap and 17--20~ms of time-of-flight\cite{supp}.  Condensate fraction for the $\Ket{0}$ atoms is measured using a two-component fit to images.  We define $N_0/N$ for the $\Ket{1}$ component as the fraction of atoms in the narrow peaks observed in time-of-flight images, measured using the procedure from Ref.~\cite{mckay:2012}.

\begin{figure}
\includegraphics[width=0.45\textwidth]{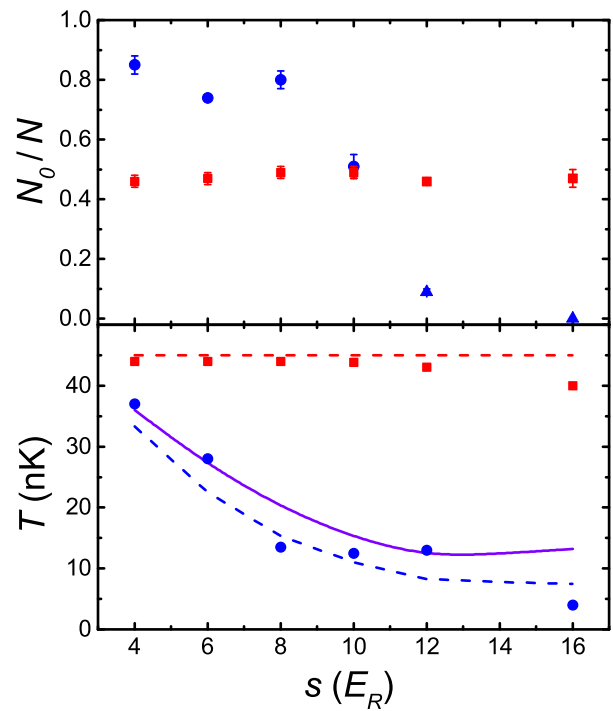}
\caption{(Color online) (Top) Condensate fraction versus lattice depth for each species. Blue (red) points are for the $\Ket{1}$ ($\Ket{0}$) atoms.  Each point is averaged over 9--15 experimental runs; the error bars represent the standard error in the mean. For data shown as circles, $N_0/N$ was measured after turning off the lattice potential in less than 50~ns. For data shown as triangles, $N_0/N$ was measured after bandmapping in 100~$\mu s$, which may overestimate condensate fraction and underestimate temperature \cite{mckay:2012}. (Bottom) Temperature inferred from the measured $N_0/N$ and number of atoms; the statistical uncertainty $T$ is less than 6\%. Mixed Mott insulator and superfluid phases exist (at zero temperature) in the lattice for $s\gtrsim13$.  For the prediction shown as dashed lines, inter-species thermalization is assumed to be absent, and entropy is conserved separately for each component.  The solid line assumes inter-species thermal equilibrium. \label{fig:loading_data}}
\end{figure}

As shown in Fig.~\ref{fig:loading_data}, $N_0/N$ for the $\Ket{1}$ atoms decreases for higher $s$, consistent with quantum depletion induced by interactions.  In contrast, $N_0/N$ is insensitive to $s$ for the $\Ket{0}$ atoms, remaining unchanged from the initial value in the dipole trap across $s=4$--16.  To interpret measured quantities such as $N_0/N$ throughout this work, we use thermodynamic calculations based on site-decoupled mean-field theory (SDMFT) \cite{sheshadri:1993,PhysRevA.78.015602} and the local density approximation for the $\Ket{1}$ atoms and the semi-ideal model \cite{naraschewski:1998} for the $\Ket{0}$ atoms.  Temperature inferred in this manner from $N_0/N$ is shown in Fig.~\ref{fig:loading_data} for both species as $s$ is varied.  Even though $T$ for the $\Ket{1}$ atoms determined in this manner has an unknown systematic error (SDMFT does not include all relevant low-energy excitations \cite{PhysRevA.84.033602}), it is still a useful measure since it is monotonically related to the actual temperature.  The temperature of the $\Ket{0}$ atoms remains fixed up to $s=16$, while the temperature of the $\Ket{1}$ atoms decreases by approximately an order of magnitude, indicating that the species are not in equilibrium.  The data show a significant deviation from the equilibrium temperature predicted assuming that the lattice turn on is isentropic (solid line in Fig.~\ref{fig:loading_data}).  At all lattice depths, the $\Ket{1}$ ($\Ket{0}$) temperature is lower (higher) than the equilibrium temperature, consistent with the lattice removing energy from the $\Ket{1}$ atoms and suppression of inter-species energy exchange. This measured deviation from the predicted equilibrium temperature for both species indicates that insignificant thermalization occurs on a timescale for which equilibration is nearly complete in the dipole trap.

To quantitatively characterize inter-species thermalization, we selectively heat the $\Ket{1}$ atoms and measure the rate of heat transfer to the $\Ket{0}$ component.  We choose this approach since it enables us to create the largest possible temperature difference and because interpreting the heat transfer rate requires knowledge of the inter-species temperature difference, which is complicated by the absence of a verified, simple method for measuring the temperature of the $\Ket{1}$ atoms at low temperature.  In fact, a primary motivation for this work is using the $\Ket{0}$ component as a thermometer \cite{mckay:2010}.  While the temperature of the $\Ket{0}$ component can be measured straightforwardly, the temperature difference between the two components cannot be controlled as the lattice potential depth is varied without knowledge of $T$ for the $\Ket{1}$ atoms.

We overcome this complication by heating the $\Ket{1}$ component to infinite kinetic temperature $\widetilde{T}$ with respect to the quasimomentum degree of freedom.  Although this may seem to be an unphysical limit, in a single-band lattice system $\widetilde{T} \rightarrow \infty$ is well defined. It does not correspond to infinite energy because the kinetic energy is bounded by $4Dt$, where $D$ is the dimensionality and $t$ is the Hubbard tunneling energy. We define the $\widetilde{T} \rightarrow \infty$ limit as $12t \ll k_BT \ll E_{bg}$, where $E_{bg}$ is the bandgap, and $k_B$ is Boltzmann's constant.  In this $\widetilde{T} \rightarrow \infty$ limit, all quasimomentum states in the lowest energy band are equally occupied.  We use the well-established procedure of dephasing \cite{greiner:2001} to generate this configuration, which involves transiently raising and lowering $s$ three times\cite{supp}. Because the $\Ket{0}$ component does not experience the lattice potential, the dephasing procedure has no direct effect on its temperature.  Images of the $\Ket{1}$ atoms after dephasing show that $\widetilde{T}\rightarrow\infty$ has been achieved and that the temperature associated with the density distribution is unaffected \cite{supp}.

To measure the heat transfer rate, we turn on the lattice over 50~ms, dephase the $\Ket{1}$ component, wait a time $t_{\mathrm{hold}}$ for heat to transfer to the $\Ket{0}$ component with the atoms held in the lattice and trap, and then measure $T$ for the $\Ket{0}$ component using time-of-flight imaging.  The rate of change of $T$ for the $\Ket{0}$ component $\dot{T}$, determined by varying $t_{\mathrm{hold}}$ at fixed $s$, is proportional to the rate of inter-species heat exchange.  Direct comparison of $\dot{T}$ between different lattice potential depths is possible since the temperature difference immediately after the dephasing step is independent of $s$.  The temperature for the $\Ket{0}$ component after loading the lattice is insensitive to $s$, and we fix $\widetilde{T}\rightarrow\infty$ for the $\Ket{1}$ component.  Furthermore, since collisions transfer only kinetic energy between species, $\dot{T}$ is solely sensitive to the kinetic energy degrees of freedom.

\begin{figure*}[htp!]
\includegraphics[width=0.95\textwidth]{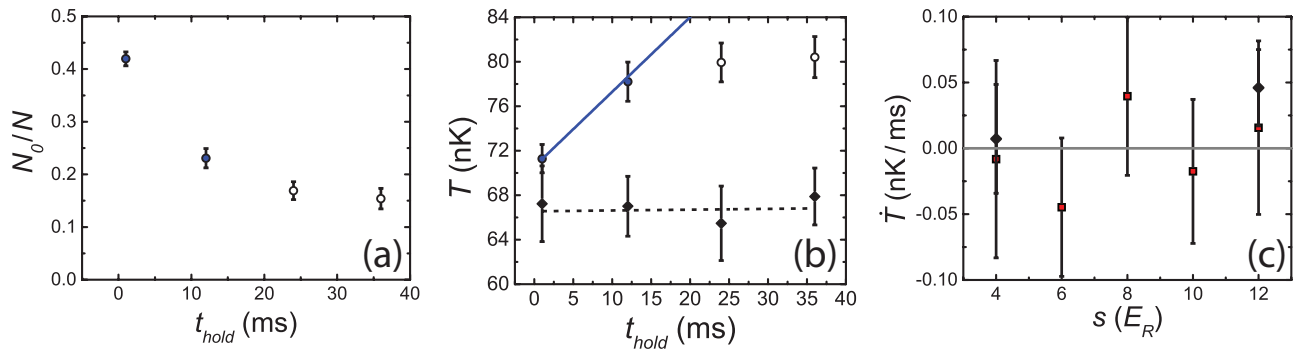}
\caption{(Color online) (a) Condensate fraction for the $\ket{0}$ component after dephasing versus hold time for $s=4$. Each point is the average of 10 experimental runs, and the error bars show the standard error of the mean. (b) Temperature inferred for the $\Ket{0}$ component from the data in (a) (circles) and from a measurement when the lattice atoms are not dephased (diamonds).  The solid line is a linear fit to data taken after dephasing and with $t_{\mathrm{hold}}=0$--12~ms (solid circles), while the dashed line is a linear fit to the control data (diamonds) for all $t_\mathrm{hold}$. (c) Control data for the heating rate. The black diamonds show $\dot{T}$ when the lattice atoms are not dephased, and the red squares show $\dot{T}$ when no lattice atoms are present. For comparison, the calculated heating rate for the $\Ket{0}$ atoms from spontaneous emission driven by the lattice laser beams is 0.02~nK/ms~$\times s\left[E_R\right]$, which is consistent with the measured heating rates at low $s$ within 1.5 times the standard error. \label{fig:dephase_data}}
\end{figure*}

Sample data for $s=4$ are shown in Fig.~\ref{fig:dephase_data}.  We observe that $T$ for the $\Ket{0}$ component increases after the dephasing step.  Because significant relaxation is measurable after $t_{\mathrm{hold}}>15$~ms at low $s$ (Fig. 3b), we define $\dot{T}$ for all $s$ as the slope of a linear fit (i.e., the short-time limit of exponential relaxation) of $T$ for the $\Ket{0}$ component for $t_{\mathrm{hold}}=0$--12~ms.  To verify this technique, we measure $\dot{T}$ at different lattice depths without the dephasing step and without the $\Ket{1}$ component present.  As shown in Fig.~\ref{fig:dephase_data}b, $\dot{T}$ is consistent with zero for these measurements at all $s$ sampled here, demonstrating that any heat transfer observed is induced by dephasing and arises from inter-species interactions.

\begin{figure}
\includegraphics[width=0.45\textwidth]{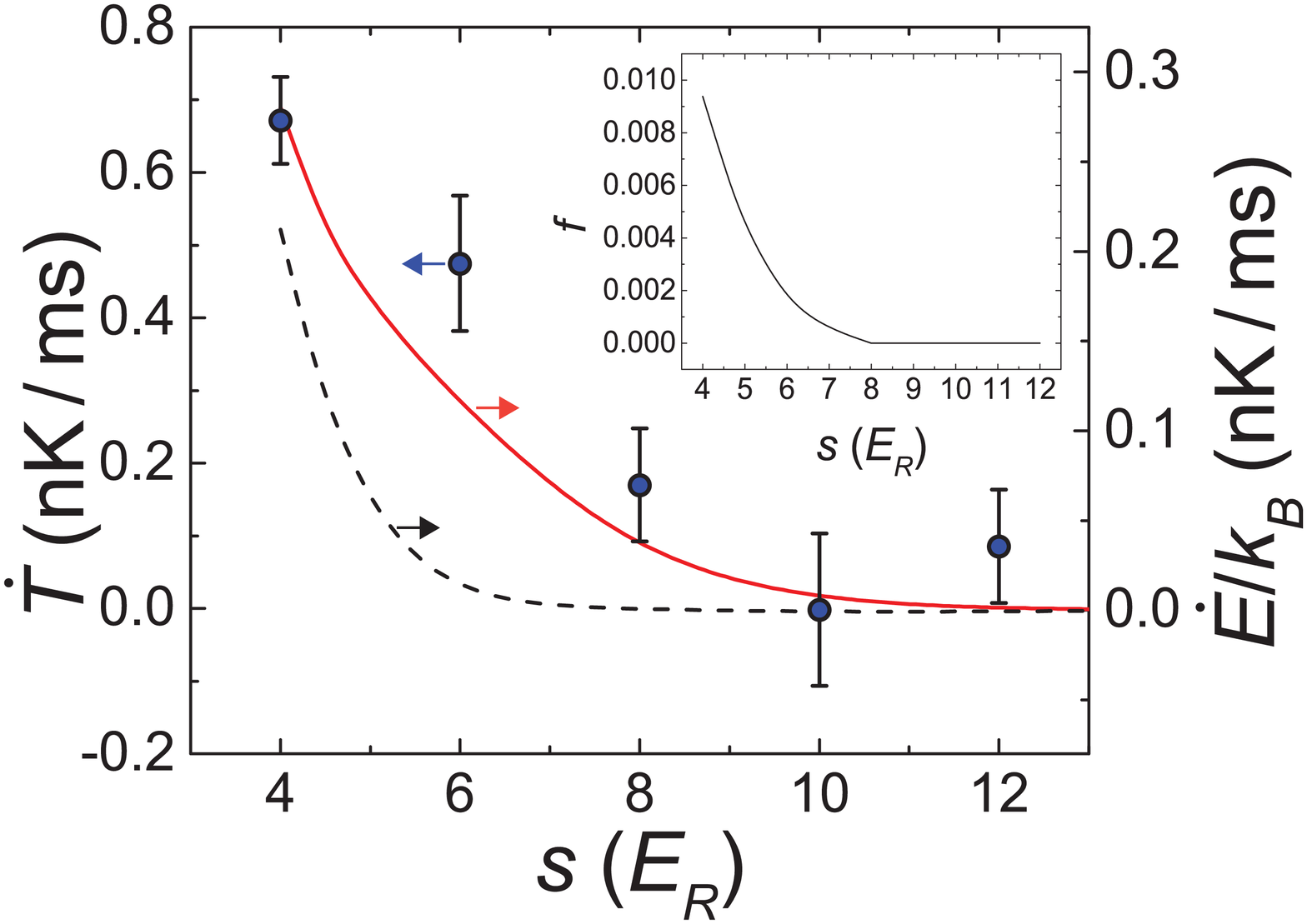}
\caption{(Color online) Energy exchange rate for different lattice potential depths, all in the superfluid regime of the BH model.  Each  point is derived from a fit to data at constant $s$, such as the solid line in Fig. 3b; the error bars are the fit uncertainty.   The lines show a FGR calculation of the energy exchange rate $\dot{E}$.  The scale has been adjusted so that $\dot{E}$ and $\dot{T}$ coincide at $s=4$ since these quantities are proportional.
\label{fig:calc_data}}
\end{figure}

As shown in Fig.~\ref{fig:calc_data}, $\dot{T}$ decreases monotonically with increasing lattice depth and is consistent with zero for $s\gtrsim10$.  Thus, the inter-species thermalization rate is suppressed as the lattice depth increases, in agreement with the disruption of equilibrium evident during turning on the lattice (Fig.~\ref{fig:loading_data}).  The measured upper bound on the thermalization rate at $s=12$ implies a thermalization time greater than 500~ms, assuming that a temperature change equal to the bandwidth must occur.  While $\dot{T}$ is expected to decrease with $s$ since the total energy added to the $\Ket{1}$ atoms during dephasing is proportional to $t$ (which approximately decreases exponentially with $s$), single-particle energetic arguments cannot explain $\dot{T}$ disappearing for $s\gtrsim10$.


We compare the measured $\dot{T}$ to a Fermi's golden rule (FGR) calculation for the inter-species energy exchange rate appropriate to the short time limit of relaxation to thermal equilibrium.  We treat both states as uniform gases occupying a volume $V$, with the $\Ket{1}$ atoms as non-interacting lattice gas at $\widetilde{T}=\infty$ and the $\Ket{0}$ atoms as a weakly interacting Bogoliubov gas at $T=0$.  Inter-species energy exchange proceeds via collisions in which a lattice particle with quasimomentum $\vec{q}$ scatters from the $\Ket{0}$ condensate at rest into $\vec{q}\,'$ and creates a $\Ket{0}$-excitation of momentum $\vec{p}$.  The rate for this process is $\Gamma = \frac{2\pi}{\hbar}\left|\left\langle \vec{q}\,',\vec{p}\left|V_{\mathrm{int}}\right|\vec{q},0\right\rangle\right|^2 \delta\left[E(\vec{q})-E(\vec{q}\,')-\epsilon(p)\right]$, where $E(\vec{q})=2t\sum_{i=1}^{3}\left[1-\cos(q_i d/\hbar)\right]$  and $\epsilon(p)=\sqrt{\left(cp\right)^2+\left(p^2/2m\right)^2}$ ($c$ is the $\Ket{0}$-component speed of sound).  The interaction $V_{\mathrm{int}}$ is the contact interaction for s-wave collisions summed over all $N^{(0)}$ $\Ket{0}$ particles. The matrix element in $\Gamma$ is $\left|\left\langle \vec{q}\,',\vec{p}\left|V_{\mathrm{int}}\right|\vec{q},0\right\rangle\right|^2  = N^{(0)} \left(\frac{4\pi a \hbar^2}{mV}\right)^2 \frac{p^2}{2m \epsilon(p)} \delta_{\vec{q},\vec{q}\,'+\vec{p}}$, where $a$ is the scattering length \cite{PhysRevA.61.063608,PhysRevLett.80.3419}. To calculate the total rate of energy transfer $\dot{E}=N^{(1)}\sum_{\vec{p},\vec{q}}\Gamma \epsilon(p) \rho_{\vec{q}}$ to the $\Ket{0}$ atoms, we sum over all final excitation momenta $\vec{p}$ and average over a uniform distribution $\rho_{\vec{q}}$ of initial quasimomenta, where $N^{(1)}$ is the number of $\Ket{1}$ atoms in the volume $V$.  Finally, we apply the local-density approximation and integrate over the trapped condensate density profile, and we approximate the $\Ket{1}$ density as one particle per site.

In qualitative agreement with the measured $\dot{T}$, the predicted $\dot{E}$ (dashed line in Fig. 4) decreases with increasing lattice potential depth and vanishes at sufficiently high $s$.  The essential ingredient in the FGR calculation that gives rise to a threshold lattice depth for thermalization is conservation of energy and momentum in collisions.  Since the $\Ket{1}$ and $\Ket{0}$ atoms have entirely different dispersion relations $E(\vec{q})$ and $\epsilon(p)$, satisfying both energy and momentum conservation can reduce the phase-space of allowed collisions \cite{griessner:2006}. The inset to Fig. 4 shows the fraction $f=\sum_{\vec{p},\vec{q}}\rho_{\vec{q}}\,\delta\left[E(\vec{q})-E(\vec{q}\,')-\epsilon(p)\right]\delta_{\vec{q},\vec{q}\,'+\vec{p}}/\sum_{\vec{q}}1$ of phase space (averaged over the $\Ket{1}$ quasimomentum distribution) for elastic collisions, which vanishes for $s\gtrsim8$.  The disappearance of $\dot{T}$ at high $s$ is evidence that dispersion mismatch prevents thermalization between the $\Ket{1}$ and $\Ket{0}$ gases.  The same mechanism is responsible for Kapitza resistance \cite{RevModPhys.41.48}, which occurs at interfacial surfaces, and can effect conductor-insulator transitions \cite{Chien2012}.

Evident in Fig. 4 is a disagreement between the observed and predicted $s$ at which the thermalization rate vanishes. To test if the strong interactions between $\Ket{1}$ atoms may be responsible for this discrepancy, we heuristically account for interactions in the FGR calculation by relaxing energy conservation.  The solid line in Fig. 4 shows $\dot{E}$ when the $\delta$-function in $\Gamma$ is replaced by a Gaussian with a standard deviation equal to the Hubbard interaction energy $U$.  The close agreement between this case and the measured energy exchange rate suggests that a full understanding of inter-species thermalization will require a more sophisticated treatment of the strong interactions induced by the lattice.

The suppression of energy exchange we observe has critical implications for schemes that rely on fast energy exchange between a lattice and a free-particle bath.  Impurity thermometry \cite{mckay:2010} and cooling methods that require equilibration \cite{ho:2009} are problematic, while techniques that rely on thermal isolation may be viable.  In particular, our measurements support the feasibility of the cooling technique proposed in Ref. \cite{griessner:2006}, which relies on one-way heat transfer to a free-state bath.  Further measurements on dynamics in species-specific optical lattices may also provide new insight into open questions regarding interfacial effects, such as Kapitza resistance.

\begin{acknowledgements}
We acknowledge Erich Mueller and Stefan Baur for suggesting the $\widetilde{T}=\infty$ technique and for initial guidance on the FGR calculation. This work was supported by the DARPA OLE program, the Army Research Office, and the National Science Foundation.
\end{acknowledgements}

\bibliography{spindeplattice_therm_vb}

\end{document}